\documentclass[12pt, preprint]{aastex}

\newcommand{\etal}   {{\rm ~et al.}}
\newcommand{\kms}    {\ifmmode{{\rm ~km~s}^{-1}}\else{~km~s$^{-1}$}\fi}

\newcommand{\tf} {\tablenotemark{(f)}}
\newcommand{\tg} {\tablenotemark{(g)}}
\newcommand{\tz} {\tablenotemark{(h)}}
\newcommand{\ti} {\tablenotemark{(i)}}
\newcommand{\tj} {\tablenotemark{(j)}}
\newcommand{\tk} {\tablenotemark{(k)}}
\newcommand{\m} {$-$}

\shortauthors{Greenhill et al.}
\shorttitle{A New H$_2$O Maser in IRAS\,F01063-8034}

\begin{document}

\title{A Search for H$_2$O Maser Emission in Southern AGN and Star Forming Galaxies -- 
Discovery of a Maser in the Edge-on Galaxy IRAS\,F01063-8034}

\author{
L. J. Greenhill,\altaffilmark{1}
S. P. Ellingsen,\altaffilmark{2}
R. P. Norris,\altaffilmark{3} 
P. J. McGregor,\altaffilmark{4}
R. G. Gough,\altaffilmark{3} \\
M. W. Sinclair,\altaffilmark{3}
D. P. Rayner,\altaffilmark{2}
C. J. Phillips,\altaffilmark{2,5} 
J. R. Herrnstein,\altaffilmark{1,6} \\
and J. M. Moran,\altaffilmark{1}
}

\altaffiltext{1}{Harvard-Smithsonian Center for Astrophysics, 60 Garden St, 
Cambridge, MA 02138 USA; greenhill@cfa.harvard.edu\,.}

\altaffiltext{2}{School of Mathematics and Physics, University of Tasmania,  
GPO Box 252-21, Hobart, Tasmania 7001 Australia.}

\altaffiltext{3}{Australia Telescope National Facility, CSIRO, Epping
Epping, NSW 2121 Australia.}

\altaffiltext{4}{Research School of Astronomy and Astrophysics, Institute of Advanced Studies,
Australia National University, Cotter Road, Weston Creek, ACT 2611, Australia.}

\altaffiltext{5}{Current address: JIVE, Postbus 2, Dwingeloo 7990 AA
The Netherlands.}

\altaffiltext{6}{Current address: Renaissance Technologies,
25 E. Loop Dr., Stony Brook, NY  11790.} 

\begin{abstract}

We report the cumulative results of five surveys for H$_2$O maser
emission at 1.35\,cm wavelength in 131 active galactic nuclei (AGNs) 
and star-forming galaxies, conducted
at the Parkes Observatory between 1993 and 1998. We detected one new maser, in
the edge-on galaxy IRAS\,F01063-8034, which exhibits a single, $\sim 0.1$ Jy
spectral feature at $4282\pm 6$\kms~(heliocentric) with an unusually large
$54\pm16$\kms~half-power full width.  The centroid velocity of the emission
increased to $4319.6\pm0.6$\kms~($38\pm2$\kms~width) over the 13 days between
discovery and confirmation of the detection.  A similarly broad linewidth and
large change in velocity has been noted for the maser in NGC\,1052, wherein jet
activity excites the emission.  Neither optical spectroscopy, radio-infrared
correlations, nor infrared colors provide compelling evidence of unusual
activity in the nucleus of IRAS\,F01063-8034.  Since the galaxy appears to be
outwardly normal at optical and infrared wavelengths, detection of an 
H$_2$O maser therein is unique.  The maser emission is evidence
that the galaxy harbors an AGN that is probably obscured by the edge-on
galactic disk.  The detection highlights the possibility that undetected
AGNs could be hidden in other relatively nearby galaxies. No other maser
emission features have been identified at velocities between 3084\kms~and
6181\kms.

\end{abstract}

\keywords{galaxies: active --- galaxies: individual (IRAS F01063-8034) ---
galaxies: Seyfert --- ISM: molecules --- masers}

\section{Introduction}

In active galactic nuclei (AGNs), H$_2$O maser emission  can occur in the
accretion disks of supermassive black holes and in shocks driven by jets
and winds.  The value of H$_2$O masers as astrophysical tracers is
illustrated by interferometric maps that outline the structure
of accretion disks with velocity resolutions $<1$\kms~and angular
resolutions $<1$ milliarsecsecond (mas).  The best example is NGC\,4258
\citep{miyoshi95}, followed by NGC\,1068 \citep{gg97}, and the Circinus
galaxy \citep{greenhillcircinus}.  Measurements of maser proper motions
and line-of-sight accelerations also make possible estimates of
geometric distance.  For NGC\,4258 the total fractional uncertainty in
distance is  $<7\%$ \citep{herrnstein99}, which may contribute to the
evaluation of systematic uncertainties in nongeometric measurement
techniques, such as the Cepheid period-luminosity relation
\citep{maoz99, mould2000}.

In contrast, for the Circinus galaxy and NGC\,1068 \citep{gallimore96},
interferometric observations show that {\it some} of the maser lines are
associated with a wind or jet, while in the galaxies NGC\,1052
\citep{claussen98} and Mrk\,348 \citep{peck01} {\it all} the  detected
emission is displaced from the central engines entirely and apparently excited
by prominent jets. These cases establish a second class of H$_2$O maser
emission in AGNs. Most notable, the spectrum of the NGC\,1052 maser exhibits a
distinctive single broad emission feature (50-100\kms), which is unlike the
complexes of narrow maser lines seen toward accretion disks.

Twenty-two AGNs have been confirmed to exhibit H$_2$O maser emission: NGC\,4945
\citep{santos79}, Circinus galaxy \citep{gw82}, NGC\,1068, NGC\,4258
\citep{chl84}, NGC\,3079 \citep{h84,hb85}, NGC\,6240 \citep{h84, hagi01},
Mrk\,1, Mrk\,1210, NGC\,1052,
NGC\,1386,  NGC\,2639, NGC\,5506, NGC\,5347, NGC\,5793,
ESO\,103-\,G\,34, IC\,1481, IC\,2560 \citep{braatz96},  IRAS\,F22265-1826
\citep{koekemoer95}, M\,51 \citep{hom51}, NGC\,3735
\citep{greenhill3735}, Mrk\,348 \citep{falcke2000}, and
IRAS\,F01063-8034 (this work).  However, detectable H$_2$O maser
emission also arises in extragalactic star-forming regions, constituting
a third class of extragalactic maser. Known sources lie in M\,33
\citep{churchwell77, huchtmeier78, huchtmeier88},  IC\,342
\citep{huchtmeier78}, M\,82 \citep{chl84}, IC10
\citep{hwb86}, NGC\,253 \citep{hom51}, NGC\,2146 \citep{tarchi01}, 
and the Magellanic Clouds
\citep{sb81, sb82, whiteoak83, wg86}. \citet{h84, chl84}, and
\citet{huchtmeier88} report several marginal detections in other galaxies.  The
apparent luminosities 
and spectral characteristics of these sources are comparable to
masers in Galactic star-forming regions.

In past searches for extragalactic H$_2$O maser emission, more than 1000 galaxies have
been observed.  Two observations suggest masers are located 
preferentially in AGNs for which the geometry of the material in the 
central parsec is
edge-on.   First, {\it all} maser-host galaxies for which X-ray observations are
available exhibit high X-ray obscuring columns $>10^{23}$ cm$^{-2}$ (e.g.,
\citet{braatz96}). Second, no masers are known in Seyfert\,1 nuclei or other
objects with relatively exposed broadline regions.  Surveys of Seyfert\,2
galaxies and low ionization nuclear emission regions (LINERs) achieve a
5-10\% detection rate for nearby objects (e.g., \citet{braatz97}) that declines
with distance, probably because of limited instrument sensitivity
(IRAS\,F22265-1826 contains the most distant known maser, with
$V_{helio}\sim7570\kms$, where we have assumed the optical definition of Doppler
shift.) 

Although high obscuring columns and edge-on geometries may describe many
maser hosts, we note that the cases of NGC\,1052 and Mrk\,348
demonstrate that detectable maser emission can sometimes be associated with
intense jet activity, and hence could in principle be observable in AGNs
whose central parsecs are not viewed edge-on.    In addition to gross geometric
effects,
the detection rates in past searches were  probably also influenced by (1) the
range of isotropic luminosity among masers, (2) chance alignments between
(amplifying) maser regions and background nonthermal continuum sources along the
line of sight, (3) warps in circumnuclear and accretion disks, (4) the survival
of quiescent, warm, molecular gas in AGN environments, and (5) anisotropic
beaming of maser radiation.

We present the results of five surveys conducted at the Parkes
Observatory between 1993 and 1998. In Section 2 we discuss the
observations and source samples. In Section 3, we discuss the discovery
of H$_2$O maser emission in IRAS\,F01063-8034, new radio
images, optical spectra, and the infrared colors  of the galaxy. In Section 4, we
conclude the discussion of whether IRAS\,F01063-8034 contains an AGN,
and the implications thereof.

\section{Parkes Survey}

\subsection{Observations}

We observed the $6_{16}-5_{23}$ transition of H$_2$O ($\lambda 1.3$ cm)
in position-switched total-power mode during five sessions between 1993
and 1998 (Table\,1) with the Parkes 64-m radio telescope of the
Australia Telescope National Facility.  For most targets, to reduce the
effects of atmospheric variability, we observed on and off-source scans
(5 minutes each) through the same patches of sky. (We followed an
off-source scan pointed $5^m$ west of the target, with two on-source
scans, and a second off-source pointing $5^m$ east of the target to
obtain a total on-source time of 10 minutes.)  For galaxies in which we
attempted to detect maser emission associated with star formation, and 
therefore observed multiple fields, we used pairs of fields as
on-off pairs whenever practical.

To maximize the number of fields we could observe during the alotted
telescope time, we collected enough data to permit a relatively coarse
flux density calibration during each session, with the intention that we
would refine that calibration following the detection of new H$_2$O
masers.  Hence, the calibration of noise levels in spectra obtained in
1993 - 1997 is only accurate to 30-50\%, depending on the epoch.   
In general, we employed
cross scans of calibrator sources to verify that pointing accuracy was
generally better than $15''$, or about 20\% of the primary beam.  When
feasible, we also estimated zenith opacities from system temperature
data (typically $\la0.1$) and corrected for the elevation dependence of
antenna sensitivity \citep{bourke94, greenhillcircinusvar}.

In 1993, we used the dual channel prime focus maser receiver and  64 MHz
($\sim 850$\kms) digital autocorrelation spectrometer, which provided
1024 channels ($\sim 0.84$\kms) in each circular polarization.  For each
AGN, we centered the two observing bands approximately on the systemic
velocity.  For each star-forming region targeted, we centered the
observing bands on the approximate  line-of-sight velocity of the local
interstellar medium, obtained from published H\,I or H$\alpha$
spectroscopy.  We grouped the observations so as to minimize the time
spent retuning the 100 MHz-wide receiver bandpass (which required the
antenna to be stowed).  We adopted a $\sim 5.7$ Jy\,K$^{-1}$ peak
telescope sensitivity, based on prior antenna calibrations, which
corresponds to an aperture efficiency of $\sim 32\%$ over the
illuminated inner 44m of the antenna.  The calibration uncertainty was
roughly 30\%.

Beginning in 1995, we used a new dual channel cryogenic high mobility 
electron transistor (HEMT) receiver
and a 500 MHz bandwidth.   We continued to use  the 1024 channel
autocorrelator and observed four contiguous 32 MHz bandpasses arranged
to cover a broader 92 MHz or $\sim 1200$\kms~bandwidth with a channel
spacing of $\sim 0.84$\kms.  Operations and calibration were conducted
as in 1993, with a roughly 30\% uncertainty in flux density calibration.

In 1996, we observed four 32 MHz bands, two in each polarization,
covering a reduced 52 MHz bandwidth, corresponding to $\sim 700$\kms,
with $\sim 0.84$\kms~channel spacing.  The narrower instantaneous
bandwidth permitted dual polarization operation that partly offset the
loss in antenna gain caused by structural deformations resulting from
installation of a new prime focus cabin. These observations were notable
because of increased calibration uncertainty (50\%) due to poor
calibration of system temperature and the unknown elevation dependence
of antenna gain associated with the new focus cabin.

In 1997 we again achieved a calibration accuracy of roughly 30\% and
estimated the peak antenna sensitivity to be 8.4 Jy\,K$^{-1}$, based on
cross scans of Virgo\,A (21 Jy).  We adopted this sensitivity to
calibrate the 1996 data.  We observed four 64 MHz bands, two in each
polarization, with $\sim 1.7$\kms~channels. However, we observed each
source twice, once with the bands offset to include largely redshifted
velocities and once to include largely blueshifted velocities.  In this
way we synthesized a broader effective bandwidth than in previous
sessions. For sources that had not been studied previously, we observed
a 199 MHz effective bandpass ($\sim
2700$\kms) centered on the systemic velocity.  For sources that had been studied previously (with
spectrometer bandwidths on the order of 50 MHz), we increased the
blue-red offset to cover effectively 244 MHz ($\sim 3300$\kms) and did
not reobserve a 45 MHz band centered on the systemic velocity.

The observations in 1998 benefitted from the use of the new Parkes
multi-beam correlator and from higher antenna gain following
adjustment of the antenna surface.  We obtained four contiguous 64 MHz
bandpasses ($\sim 0.84$\kms~channels) that we spread out to provide an
instantaneous bandwidth of 236 MHz ($\sim 3100$\kms).  Using
observations of Virgo\,A, for which we adopted a flux density of 21 Jy
and a 10\% correction due to partial resolution of the source
\citep{kuiper87},  we measured a 6.3 Jy\,K$^{-1}$ antenna sensitivity. 
We corrected for atmospheric opacity and achieved a 20-30\% calibration
uncertainty overall.

\subsection{Source samples}

We employed several different source samples in the five Parkes surveys,
including star forming galaxies, optically identified AGNs, and obscured
AGNs.  In 1993, we selected southern galactic nuclei ($\delta<-20^\circ$)
with $100\mu$m IRAS flux densities $>20$ Jy.  We also observed
extragalactic star forming regions in the Sculptor and Centaurus A
groups, the NGC~2997 and NGC~6300 associations, the Local Group, and
several field galaxies \citep{devaucouleurs75, hg82, kt79}.  The
observations of southern Local Group galaxies complemented an earlier study
of northern Local Group members \citep{greenhill90}.    When we
formulated the survey, nine of the eleven known extragalactic H$_2$O
masers were associated with the 83 known IRAS galaxies with $100\mu$m
flux density $>50$ Jy.  However, in retrospect, discovery of these nine
masers may have depended more on proximity of the galaxies to the Sun than on a (hoped for) direct
physical relationship between maser emission (which arises in parsec scale
structures) and IRAS far-infrared emission (which also originates on scales
that are orders of magnitude larger).  Among the larger number of H$_2$O
masers known today, there is no apparent correlation.  Galaxies with similar IRAS
$100\mu$m flux densities can have peak maser flux densities that differ by over
an order of magnitude, and visa versa. 

In 1995, we selected objects from two samples of active galaxies:  (1)
southern ``radio-excess'' IRAS galaxies \citep{rn97} and (2) nearby
galaxies ($z<0.09$; $\delta < +20^\circ$) that harbor hard X-ray sources
($E > 2$ keV) toward which substantial X-ray absorbing columns had been
observered, using data from the EXOSAT \citep{t88}, HEAO-1
\citep{weaver95}, GINGA \citep{awaki}, and ASCA satellites (R. Mushotzky 
1995, private communication).   With respect to the first sample,  radio
emission that surpasses the radio-far infrared relation for normal
galaxies is an indicator of activity (e.g., Condon, Anderson, \&
Broderick 1995; see Section 4) and can substitute for conventional
optical identification when internal reddening is substantial.  
\citet{koekemoer95} observed 25 galaxies from a similar sample of
northern radio-excess galaxies and detected the maser in
IRAS\,F22265-1826.  The second sample contained AGNs for which the
measured absorbing columns were suggestive of the approximately edge-on
geometries that seem to be a prerequisite for visible maser emission from
accretion disks (see \citet{braatz96}).

In 1996, we observed galaxies detected both by IRAS and HEAO-1 (0.25-25
keV) and identified optically as Seyfert\,2 objects by \citet{ks90}.  We
also selected targets from a sample of early-type galaxies whose compact
radio cores (1-10 pc diameter) suggested the presence of AGNs. At least
one quarter of the galaxies in this sample had been shown to be
Seyfert\,2 galaxies or LINERs \citep{slee94}.

In 1997, we again concentrated on Seyfert\,2 galaxies and LINERs
($<9500$\kms) drawn from the CfA redshift survey (J. P. Huchra 1997, 
private communication), but searched specifically for high-velocity maser
emission that previous surveys of southern galaxies could not have
detected.  The emphasis on bandwidth was motivated by the observation
that in all galaxies with recognizable high and low-velocity maser
emission (except  NGC\,4258), the high-velocity emission is stronger
than the low-velocity emission, which may place narrow-band surveys 
at a relative disadvantage.

In 1998, we extended the high-velocity survey to galaxies identified by
a cross-referencing of the Parkes-MIT-NRAO (PMN) catalog \citep{pmn}
with the IRAS Faint Source Catalog \citep{moshir92}.  These galaxies had
known redshifts ($z< 0.085$) but  in many cases had at best ambiguous
spectroscopic identifications, because of substantial internal obscuration
by gas and dust.  However, as in our previous sample, the radio excesses
were suggestive of nuclear activity.  (We also included radio-excess
objects ($\delta< 20^\circ$) taken from the \citet{condon95} sample that
had not been investigated previously as possible sources of maser
emission.)

\section{Detection of an H$_2$O Maser in IRAS\,F01063-8034} 

We observed 131 AGNs (Table\,2) and detected one new H$_2$O maser, in the
edge-on galaxy IRAS\,F01063-8034 (Figure\,1).  \citet{braatz97} first
noted a weak preference for observable maser emission in highly-inclined
galaxies.  With the detection of a maser in IRAS\,F01063-8034, 
19 masers are known to lie in spiral galaxies, of which 9 have galactic
inclinations $\ga70^\circ$. We have followed-up the detection by
obtaining a confirming maser spectrum, radio images, and optical spectra.

\subsection{The H$_2$O Maser Spectrum}

We detected the H$_2$O maser with the Parkes antenna on 1998 August 27
and confirmed the detection on 1998 September 8 with the 
70-m antenna of the NASA Canberra Deep Space Communications Complex at
Tidbinbilla, Australia, which is several times more sensitive than Parkes
(Figure\,2).  At Tidbinbilla, we used a cooled, wide-band HEMT
receiver and a single polarization, 20 MHz bandwidth (270\kms), 16384
channel correlation spectrometer in position-switched total-power mode. 
We convolved the spectra with a 16 channel wide boxcar function to obtain
a 0.26\kms~channel spacing. In order to search for emission more than
$\sim 10$ MHz away from the line we observed first, we tuned the receiving
system to several band-center velocities and covered the range 4100 to 4850\kms,
achieving an RMS noise level of
$\sim 30$ mJy in each spectral channel.  (The  system
temperature changed by 30-40\% from day to day possibly because of 
weather conditions. We conservatively estimate that the calibration is
uncertain by perhaps 50\%.)

The line profile of the emission observed at Parkes is well fitted by a
Gaussian model with a $0.09\pm0.02$ Jy peak, $4282\pm6$\kms~centroid, and
$54\pm16$\kms~half-power full width (Figure\,2).  A model fitted to the
emission observed at Tidbinbilla has a $0.113\pm0.004$ Jy peak,
$4319.6\pm0.6$~centroid, and $38\pm2$\kms~half-power full width (where we
quote statistical errors).  The velocity shift of 38\kms~over 13 days is
almost unprecedented.  The only other similar occurence has been in NGC\,1052,
where the single emission feature jumped by 45\kms~between two
observations 5 months apart \citep{braatz96}.

We have confirmed the calibration of the velocity scales for our Parkes and
Tidbinbilla spectra separately at the $<1\kms$~level.  For the Parkes
data, we have compared the measured heliocentric velocities of two widely
separated lines in a spectrum of the H$_2$O maser in the Circinus Galaxy,
obtained during the 1998 August observations reported here, to an independent
spectrum created from VLBI data recorded in 1998 June.  For the Tidbinbilla
data, we have compared the centroid velocity for a persistent, isolated spectral
feature of the Mrk\,1210 H$_2$O maser measured on 1997 November 15 with that
measured with antennas at Effelsberg and
GreenBank between 1993 and 1996 \citep{braatzphd, braatz96}.  
Ultimately, the velocity scale for both stations is derived directly from the
observed band-center sky frequency determined by the tuning of receiver
elements.  At Tidbinbilla in particular, this is directly set by the observer,
and hence, the spectrometer calibration is as robust as that of the more widely
tested Parkes system. 

The mechanism responsible for the distinctive broad, centrally-peaked H$_2$O
maser lines associated with strong jet activity (i.e., NGC\,1052
and Mrk\,348) is not well understood. We speculate that the lines may
be composites of narrower, spatially distinct features. Such composites
are seen toward some AGN of late-type systems, for which 
spectral features narrower
than a few \kms~sometimes blend together to form broad but irregular complexes
(e.g., NGC\,1068).  Where maser excitation is tied to jets,
individual spectral components could correspond to individual shocks in
entrained or ambient material.  However, the largely smooth, centrally peaked
maser line profiles of these systems contrast with the irregular
appearance of maser line complexes (more commonly) associated with AGNs. 
We suggest that the contrast could reflect the difference between maser
amplification in accretion disks, which have well ordered
dynamics, and amplification in jet entrained material, which 
has relatively chaotic dynamics.
Furthermore, we note that a third example of broad maser
emission, IRAS\,F22265-1826, has been resolved into at least four
clumps distributed over $\sim 1$ pc with individual line widths of tens of
\kms~\citep{greenhill22265}, although association of this maser with an
underlying jet is less certain than it is for NGC\,1052 and Mrk\,348.

The 45\kms~shift in the maser line velocity of NGC\,1052 (in $<5$ months)
is interesting. We argue that such variation could be typical in 
masers driven by jet-activity, for which intrinsically fast fluctuations in a
(relativistic) jet, or in its interaction with the surrounding medium, may cause
new centers of emission to rise, others to decline, and the composite line to
shift in velocity. The maximum possible shift would at least
depend on the velocity dispersion of the entrained material or bulk flows
in ambient material.  Although the apparent persistence of a
smooth, centrally peaked line profile in NGC\,1052 should constrain this model,
at present, there is insufficient data from spectroscopic monitoring to do 
so.

We suggest that IRAS\,F01063-8034 contains a previously undetected AGN 
because of the broad maser line profile and unusual variation in velocity,
which are also observed in NGC\,1052.
Abrupt changes in velocity are difficult to understand in the
context of two alternate models, maser emission from a thin accretion disk, as
in NGC\,4258, or from a starburst, as in NGC\,253 and M\,82.  Accretion disks
that support maser emission are slowly varying structures, and the
apparent luminosities of known masers that are associated with star formation
are relatively small.  (The flux density of the M\,82 maser, scaled to
a nominal distance of 57 Mpc, assuming $cz\sim4300$\kms~and 
$H_\circ=75$\kms\,Mpc$^{-1}$, is $\sim 300$ times weaker than IRAS\,F01063-8034.)

\subsection{Optical Spectroscopy}

We obtained optical spectra of IRAS F01063-8034 with the Double Beam
Spectrograph \citep{rodgers88} on the Australia National University 2.3
m telescope at Siding Spring Observatory on  2000 October 1. The $4''$
slit was aligned with the minor axis of the galaxy. The B600 and R600
gratings (in the blue and red arms of the spectrograph, respectively)
provided velocity resolutions of $\sim 300$\kms~from 3800--5350 \AA\ 
and $\sim 200$\kms~from 5700--7500 \AA.  A Ne--Ar arc lamp spectrum
obtained 12 hr prior to the astronomical observation provided calibration of
the wavelength scale. Analysis of OH emission in the extracted sky
spectrum permitted some refinement of the velocity calibration 
for the red spectrum \citep{ost96};
after calibration with 22 sky emission lines, uncertainty in the
wavelength calibration was 0.37 \AA\ or $\sim$ 17 km~s$^{-1}$
($1\sigma$).  The galaxy spectra were flux calibrated using a spectrum
of EG\,131 obtained on the same night and the absolute flux calibration
of \citet{bessel99}. Figure\,3 shows the blue and red spectra for the
central $14\rlap{.}''4$ (16 pixels) along the slit.

The continuum light of late-type stars dominates the minor axis optical 
spectrum of IRAS\,F01063-8034 (Figure\,3).  The prominent central dust
lane (Figure\,1) almost  completely obscures the disk plane so most of
these stars must belong to the central bulge population. Weak emission is seen
near H$\alpha$/[N\,II], but no other emission lines are apparent, which makes
optical spectroscopic identification of nuclear activity difficult.

Based on a detailed analysis of the spectrum (Figure\,4) we suggest that
the observed emission line is  [N~II] $\lambda$6583, in which case the
implied heliocentric systemic velocity is $4285\pm35$\kms. The velocity
inferred from Na\,I\,D and Mg\,I\,b $\lambda$5183 absorption lines are
$4246\pm50$\kms~and $\sim 4140$\kms, respectively, though we emphasize
that the zero point of the blue spectrum is uncertain, because there were
not enough sky emission lines to refine the wavelength calibration.  The
three line velocities agree reasonably well with each other, which
precludes identification of the emission line as H$\alpha$, for which the
implied redshift would be $\sim 1050$\kms.   Prior estimates of the
systemic velocity are $4249\pm 27$\kms~\citep{dacosta91} and
$5047\pm21$\kms~\citep{sadler84},  both of which rely on cross
correlation of stellar absorption features.   (Dacosta et al. may
have also modelled the lone emission line reported here but it would
contribute relatively little to their published weighted mean velocity.  We
note that \citet{phillips86} report an early failure to detect [N\,II]
emission.)

The probable identification of [N\,II] emission provides some additional
evidence for there being an AGN in IRAS\,F01063-8034, specifically a
LINER.  The [N\,II] line flux is $3.4\pm0.3 \times 10^{-15}$
erg\,s$^{-1}$\,cm$^{-2}$ (comparable to the \citet{phillips86} upper
limit), and the $3\sigma$ upper limit on H$\alpha$ emission is $\sim 1.2
\times 10^{-15}$ erg\,s$^{-1}$\,cm$^{-2}$, from which we estimate that
the familiar AGN emission line ratio diagnostic
$\log$([N\,II]$\lambda$6583/H$\alpha$) is $\ga 0.44$.  This is
characteristic of LINERs among early-type galaxies \citep{VO87,
phillips86}.  Although the H$\alpha$ emission flux may be artificially low
because of stellar H$\alpha$ absorption, the effect is likely to be small
because there is only marginal evidence for H$\beta$ absorption 
(Figure\,3) and no indication of higher Balmer lines in absorption.  

\subsection{Radio Images}

We made radio snapshots of IRAS\,F01063-8034 with the Australia
Telescope Compact Array (ATCA) at 3.5\,cm and 6.3\,cm wavelength in 1998
January and February with the 6C and 6B configurations, respectively (18
minutes on-source in each of two 12 hour tracks).  We calibrated the
phase data with respect to the calibrator 0252-712 and the  amplitude
data with respect to 1934-638, for which we adopted flux densities of
2.8 Jy at 3.5\,cm and 5.8 Jy at 6.3\,cm.  The angular resolutions were
$\sim 1''$ and $\sim 2''$ at the two wavelengths, respectively.

At each wavelength, emission is limited to a weak source associated with the
galactic nucleus (Figure\,1; Table\,3).  The observed flux density at 6.3\,cm
is significantly smaller than the $81\pm7$ mJy reported in the PMN catalog
\citep{pmn}. However, this is at least in part the result of confusion in the
Parkes $>1'$ beam, because there is a second  source at about
$\alpha_{2000}=01^h07^m 04\rlap{.}^s7$, $\delta_{2000}= -80^\circ17'04''$. 
Its flux density is $\sim 13$ mJy and $\sim 27$ mJy at 3.5 and 6.3\,cm,
respectively.

We included IRAS\,F01063-8034 in our survey because its 6.3\,cm radio
flux density in the PMN catalog \citep{pmn} exceeded that expected based on its
IRAS 60 $\mu$m flux density, which suggested that the radio emission was
powered by an AGN. However, the weaker observed radio flux density isolated to
the nucleus begs the question of whether there is really any radio-excess (and
whether the galaxy should have been included in our surveys - twice).  It is
clear from Figure\,5 that IRAS\,F01063-8034 actually lies on the well known
radio-FIR correlation for starburst galaxies and radio-quiet AGNs, which
underscores that there is no radio-excess and no further evidence of
activity.

\subsection{Infrared Colors}

AGNs have been identified with some success by analyses of infrared colors.
Nuclear activity heats dust that creates excess 12 and 25 $\mu$m emission and
warm IRAS 25--60 $\mu$m colors \citep{low88, degrijp87, degrijp92}.  For
IRAS\,F01063-8034, $\log(F_{60\mu m}/F_{25\mu m}) = 1.023$, which is much
cooler than the 0.6 upper limit used by \citet{low88} to identify (warm)
candidate AGNs. \citet{rush93} used the IRAS Faint Source Catalog
\citep{moshir92} to investigate the far-infrared colors of their Extended 12
Micron Galaxy Sample. On a two color diagram IRAS\,F01063-8034 lies at the
extreme cool end of the locus of non-Seyfert galaxies (Figure\, 6), indicating
that its IRAS flux densities are dominated by cool dust typical of normal
galaxies; there is no indication of the warm far-infrared emission typical of
many Seyfert galaxies. While some Seyfert galaxies do occupy this same region
of the far-infrared two-color diagram, these are likely to be objects in which
star formation in the host galaxy dominates the far-infrared emission from the
Seyfert nucleus. The same may be true for IRAS\,F01063-8034 where we know from
the optical spectrum that any nuclear activity could be weak or highly
obscured at optical wavelengths, probably by the edge-on galactic 
dust disk. 

\section{Conclusion}

No normal galactic nuclei have been known to host H$_2$O maser emission, yet
IRAS\,F01063-8034 {\it appears} to be a normal galaxy. High excitation
optical lines are absent and common measures that rely on
estimates of nuclear far-infrared or radio continuum flux densities do
not betray substantial activity. However, variable H$_2$O maser emission
that we speculate is excited by jet activity (as in NGC\,1052 and
Mrk\,348) and the putative [N\,II] emission, are suggestive of nuclear
activity in IRAS\,F01063-8034. We propose that this galaxy contains a
heavily obscured AGN, and that it may represent a larger class of AGNs
undetectable in optical and infrared surveys, because of the inclinations of
galactic disks or smaller scale optically thick structures.  X-ray spectroscopic
observations may resolve the nature of the IRAS\,F01063-8034 nucleus, but at
present, it is the only AGN identified first by detection of H$_2$O maser
emission.  Apparently normal galaxies that are highly inclined may be profitable
targets for new surveys intended to detect extragalactic H$_2$O maser emission.

\acknowledgements
We thank Harry Fagg, Euan Troup, Warwick Wilson, and the Parkes engineering
staff for their dedication and assistance during the 5 years of
observation, in particular with the maser receiver system, correlator,
and signal processing electronics.  Paul Harbison and Edward King assisted
with collection and reduction of Tidbinbilla data. We thank Michael Bessell
for obtaining the optical spectrum and both Michael Dopita and Jim Braatz for
informative  discussions.  We also acknowledge the public service of Jim Braatz
who has maintained a cumulative catalog of galaxies that have been observed in
searches for new H$_2$O maser sources 
(http://www.gb.nrao.edu/$\sim$jbraatz/H2O-list.html).   
We are  grateful to Lee Simons for
substantial assistance in constructing Table  3 and cross-checking the contents 
with available databases. SPE acknowledges financial support from the Australian
Research Council.  This research has made use of the NASA/IPAC Extragalactic
Database (NED) which is  operated by the Jet Propulsion Laboratory, California
Institute of Technology,  under contract with the National Aeronautics and Space
Administration.


\begin{figure}[ht]
\plotone{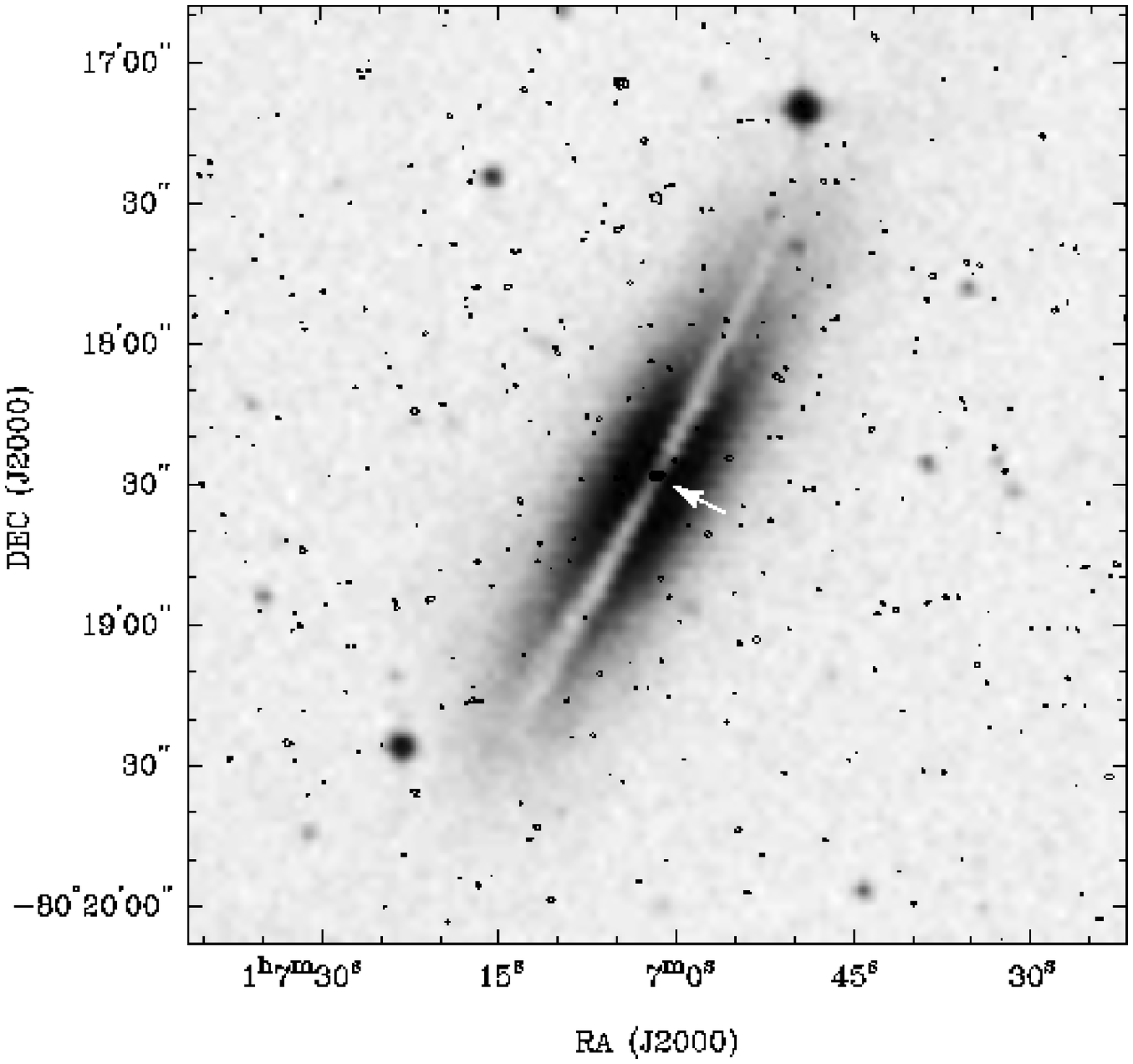}
\caption{Digitized Sky Survey blue image of IRAS\,F01063-8034 ({\it
gray scale}) with $1''$ resolution and 4 cm wavelength continuum ATCA
snapshot superposed ({\it contours}).  The arrow indicates the contours
that mark the nuclear radio source. 
}
\end{figure}

\begin{figure}[ht]
\plotone{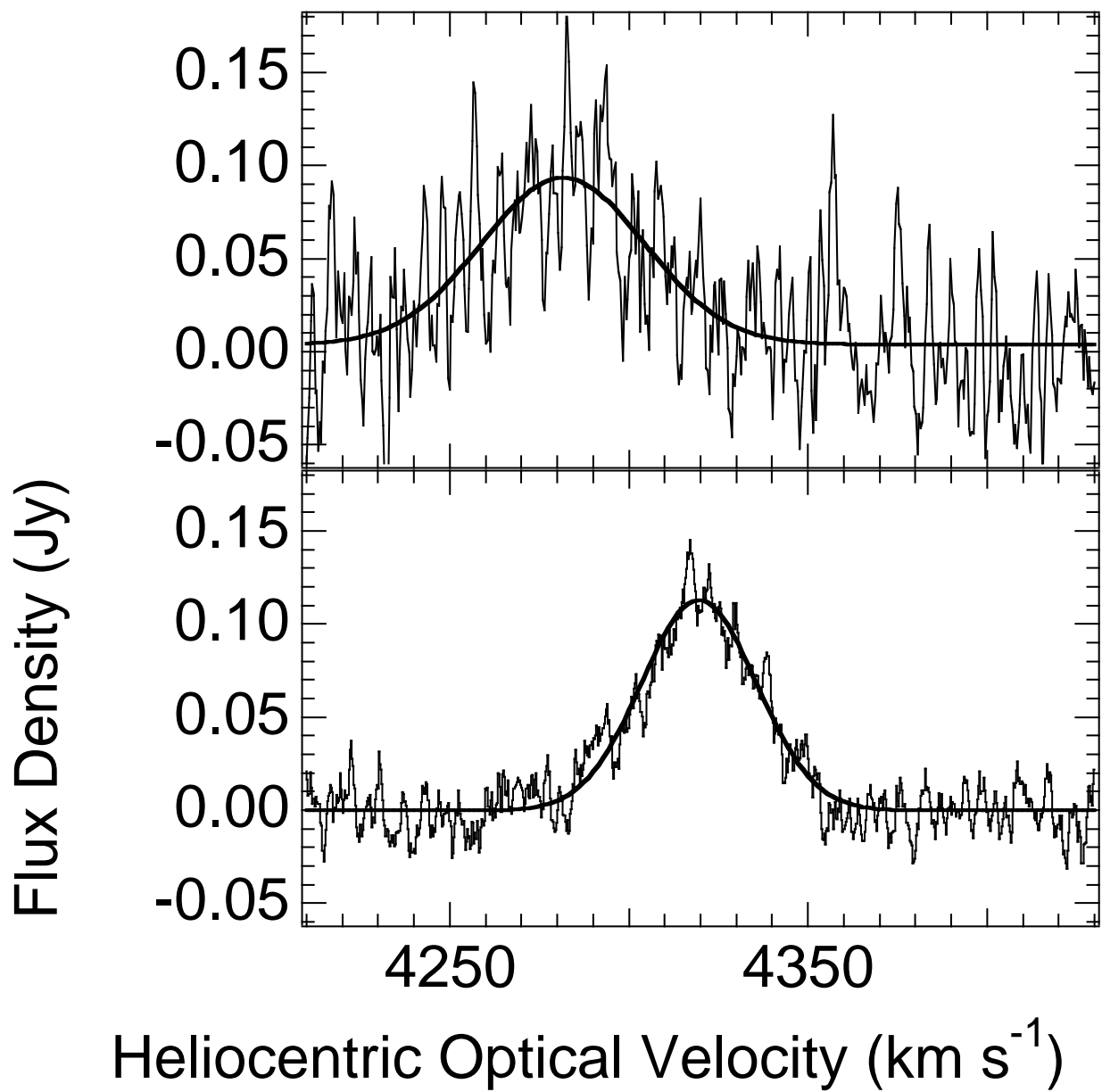}
\caption{Spectra of H$_2$O maser emission in IRAS\,F01063-8034.  {\it (top)}--
Discovery spectrum obtained with the Parkes telescope, convolved with a
2.1\kms~wide boxcar function.   {\it (bottom)}-- Confirming spectrum obtained
with the Tidbinbilla antenna 13 days later, convolved with a 1.9\kms~wide
boxcar function. The curves are the fitted Gaussian line profiles whose 
parameters are listed in the text.}
\end{figure}

\begin{figure}[ht]
\plotone{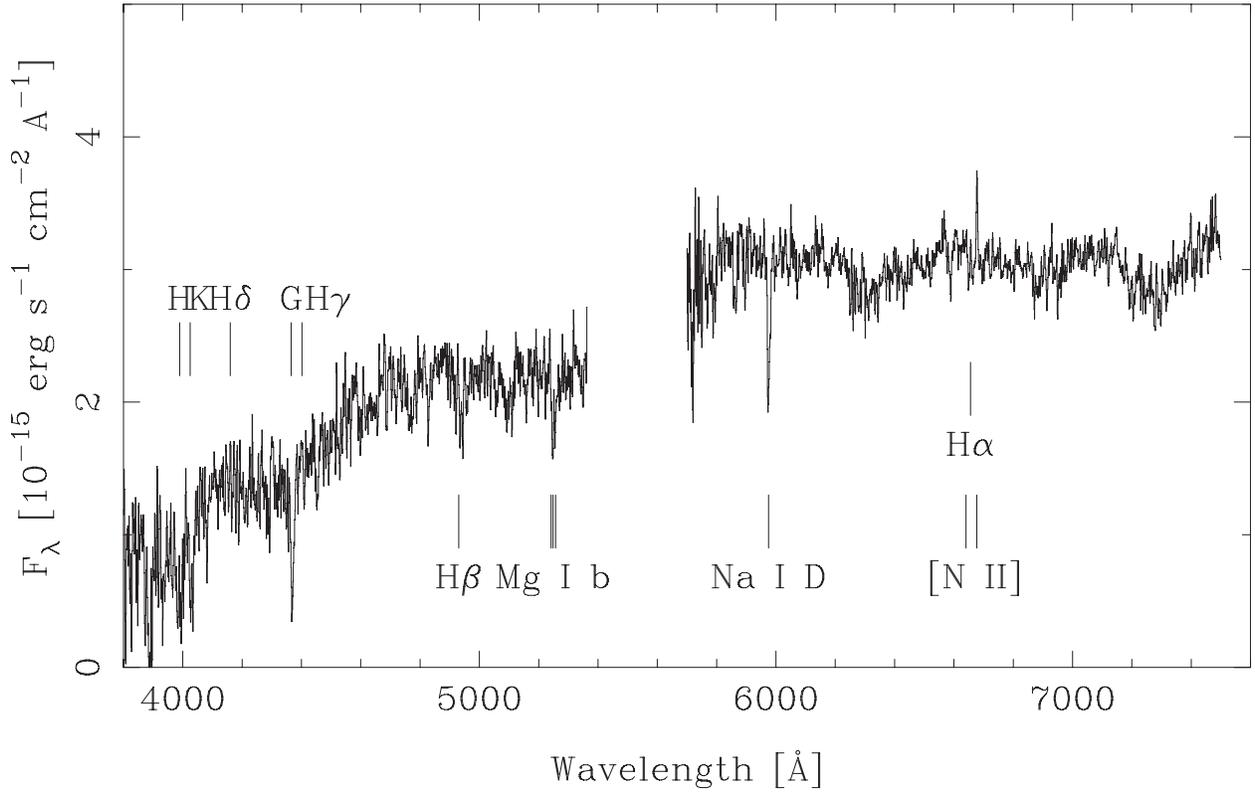}
\caption{Blue and red optical spectra of IRAS\,F01063-8034. The  vertical bars
indicate the rest wavelengths of the Ca\,II H and K, G band, Mg\,I\,b, and
Na\,I\,D absorption lines, as well as the H$\alpha$/[N\,II], and higher Balmer
emission lines.}
\end{figure}

\begin{figure}[ht]
\plotone{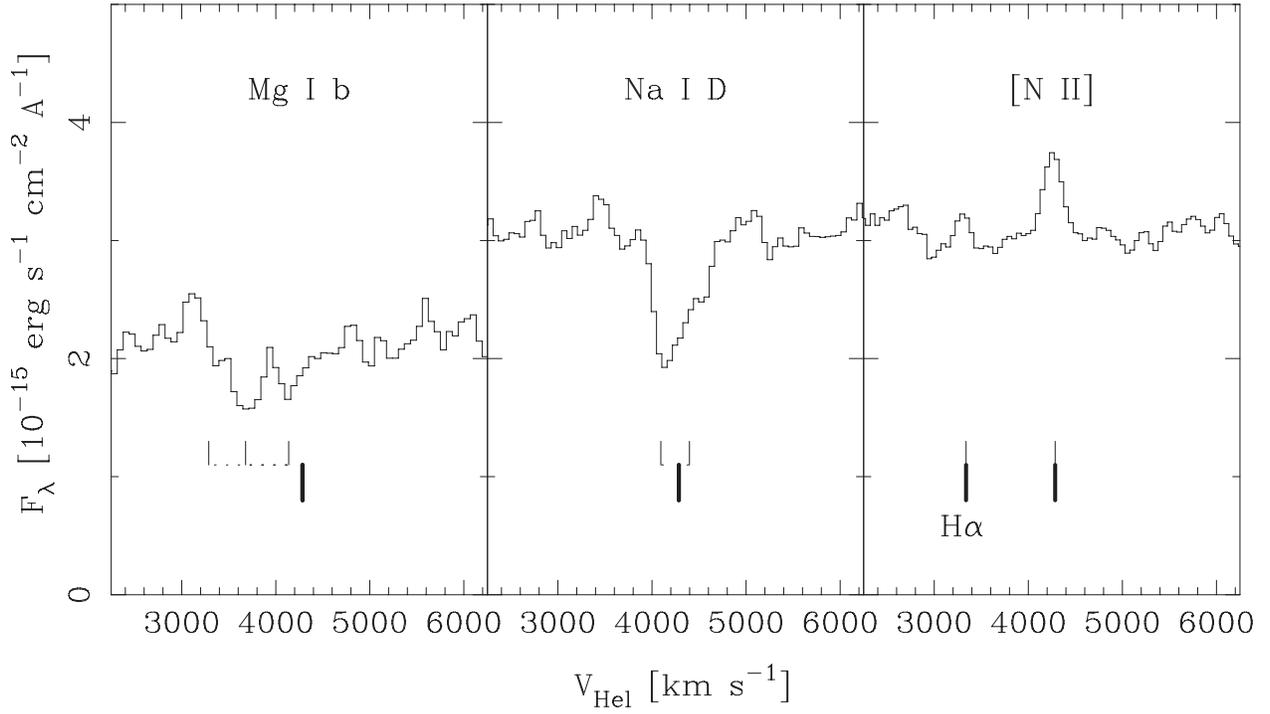}
\caption{Line profiles of the Mg~I~b $\lambda$5183.3  and Na~I~D absorption
lines, and the [N~II] $\lambda$6583 emission line, plotted with respect to
heliocentric velocity.  Heavy vertical lines in each frame indicate a velocity
of 4246\kms.  Light vertical lines denote spectral features of Mg\,I\,b
$\lambda\lambda$5168,5175,5183, Na\,I\,D $\lambda\lambda$5889,5895, and
H$\alpha$/[N~II] $\lambda$6583.}
\end{figure}

\begin{figure}[ht]
\plotone{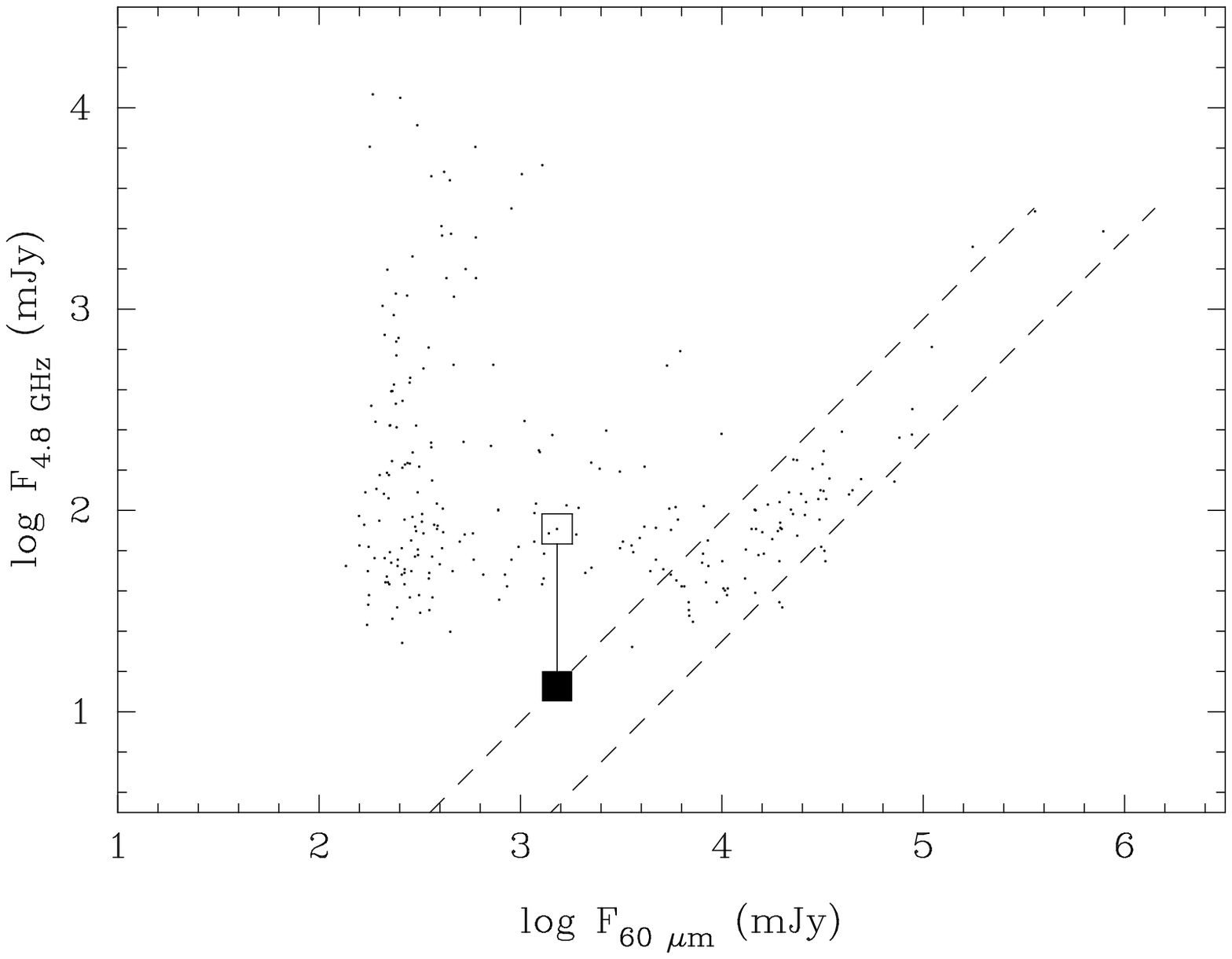}
\caption{Radio-FIR diagram for sources found by correlating the 6.3\,cm
wavelength PMN catalog \citep{pmn} with the IRAS 60$\mu$m Faint Source Catalog
\citep{moshir92}. An {\it open square} indicates the position of IRAS
F01063-8034 based on its PMN radio flux density. A {\it filled square}
indicates the revised position based on the ATCA-measured radio flux density.
{\it Dashed} lines represent the radio-FIR correlation for starbursts galaxies
and radio-quiet AGNs.}
\end{figure}

\begin{figure}[ht]
\plotone{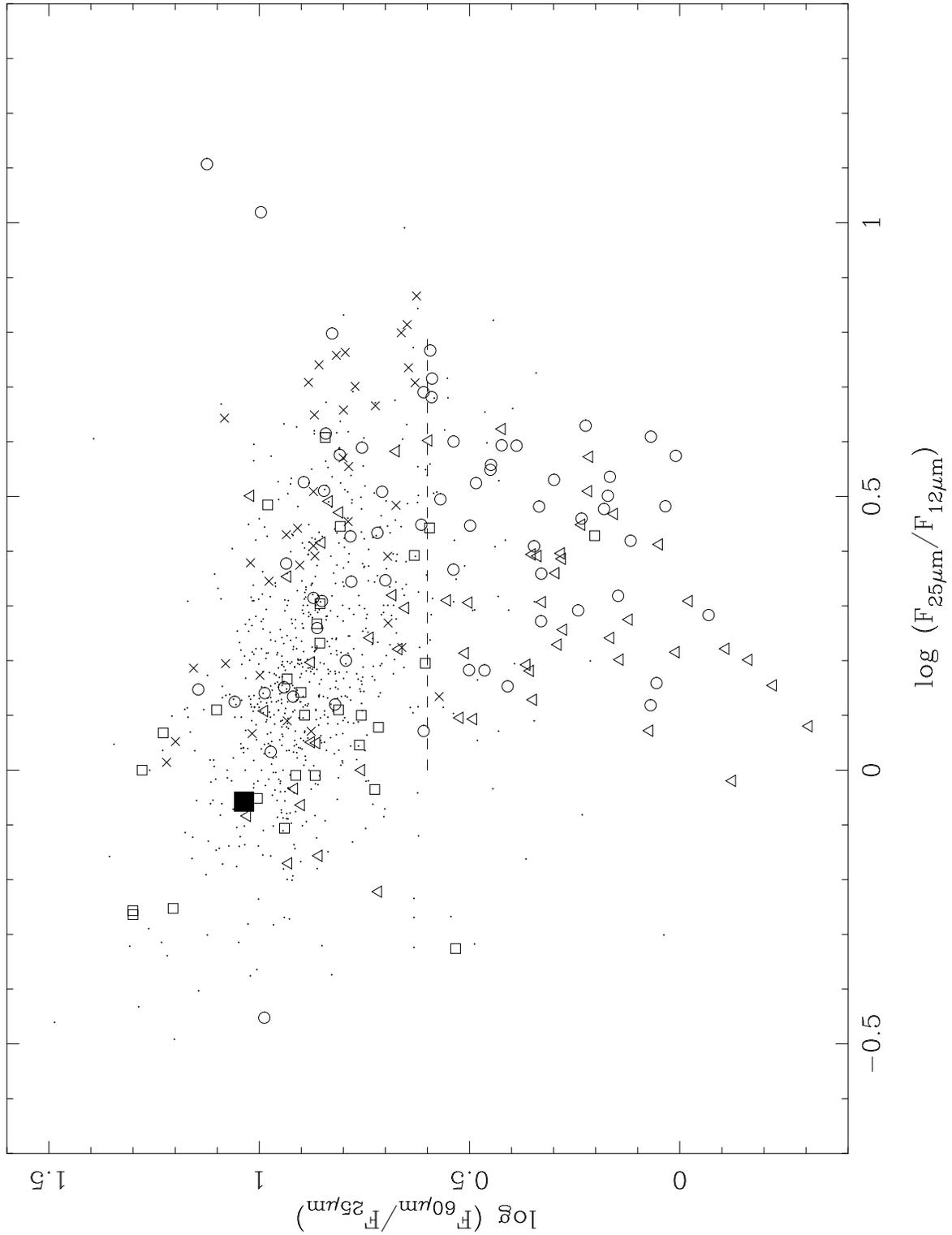}
\caption{Far-infrared two color diagram based on data for the Extended 12
Micron Galaxy Sample of Rush et al. (1993). Seyfert\,1 galaxies ({\it triangles}),
Seyfert 2 galaxies ({\it circles}), LINER galaxies ({\it squares}), 
starburst galaxies ({\it crosses}), and normal galaxies ({\it points})
are indicated. The criterion of \citet{low88} for ``warm extragalactic
objects'' ($\log (F_{60\mu m}/F_{25\mu m}) < 0.6$ is indicated by a {\it
dashed line}. IRAS F01063-8034 is shown as a {\it filled square}.}
\end{figure}

\clearpage

\begin{deluxetable}{lcccc}
\tablenum{1}
\tablecaption{Observation Log}
\tablewidth{6.0in}
\tablecolumns{5}
\tablehead{
\colhead{Dates}     &  
\colhead{Receiver}  & 
\colhead{T$_{\rm sys}$} &
\colhead{Sensitivity\tablenotemark{(a)}} &
\colhead{Calibration Unc.\tablenotemark{(b)}} \\
\colhead{}     &
\colhead{}     &  
\colhead{(K)}  &
\colhead{~~(Jy\,K$^{-1}$)} &
\colhead{(\%)} 

}      

\startdata


1993 March 11-18     &  Maser & 70-110 & 5.7 & 30           \\  
1995 October 11-16   &  HEMT  & 70-250\tablenotemark{(c)} & 5.7 & 30  \\
1996 September 11-15 &  HEMT  & 100   \tablenotemark{(d)} & 8.5 & 50  \\ 
1997 July 21-29      &  HEMT  & 100    & 8.4  & 30          \\   
1998 August 24-28    &  HEMT  & 110    & 6.3  & 20-30       \\   

\enddata

\tablenotetext{(a)}{Sensitivity of the illuminated central 44m of the antenna. 
The sensitivities estimated in 1993 and 
1997 were  adopted to calibrate the data obtained in 1995 and 1996, respectively.
The sensitivity changed significantly between 1995 and 1996
and between 1997 and 1998 because of antenna modifications.}

\tablenotetext{(b)}{Overall uncertainty in flux density calibration.}

\tablenotetext{(c)}{System temperature of prototype receiver increased with redshift. 
The maximum corresponds to sky frequencies $\sim 21$ GHz.}

\tablenotetext{(d)}{T$_{\rm sys}$ uncertain by 50\%.}

\end{deluxetable}

\clearpage

\begin{deluxetable}{llllclllrlrc}
\rotate
\tabletypesize{\scriptsize}
\tablenum{2}
\tablecolumns{12}
\tablewidth{0pt}
\tablecaption{Observations at Parkes}
\tablehead{\colhead{Source} & 
           \colhead{Alias}  & 
           \colhead{RA(1950)} & 
           \colhead{Dec(1950)\tablenotemark{(a)}} &
           \colhead{Morphology\tablenotemark{(b)}} &
      \multicolumn{2}{c}{V$_{\rm sys}$\tablenotemark{(c)}} & 
	     \multicolumn{2}{c}{Velocity Range\tablenotemark{(d)}} & 
	     \multicolumn{2}{c}{RMS Range\tablenotemark{(e)}} &
           \colhead{Epoch}                                       \\
      &&&&&\multicolumn{2}{c}{(km$\,$s$^{-1}$)} &
           \multicolumn{2}{c}{(km$\,$s$^{-1}$)} &
           \multicolumn{2}{c}{(Jy)} &
          }

\startdata
NGC\,55\tf          & \nodata           & 00 12 38.00 &\m39 29 54.0 & SB(s)m: sp                 & 129  &\hspace{-0.15in}$\pm3$   &\m302  & 561   & 0.11  & 0.13   & 1993  \\
IRAS\,F00198-7926   & \nodata           & 00 19 55.6  &\m79 26 52   & pec Sy2                    & 21825& 30  & 20013	& 22536 & 0.075 & 0.090  & 1997  \\
IRAS\,F00198-7926   & \nodata           & 00 19 55.60 &\m79 26 52   & pec Sy2                    &21825 & 30  & 21180 & 22472 & 0.057 & 0.10   & 1995  \\
IRAS\,F00344-3349   & ESO\,350-IG038    & 00 34 25.67 &\m33 49 49.0 & triple                     & 6154 & 67  & 5570  & 6739  & 0.068 & 0.074  & 1995  \\
IRAS\,F00494-3056   & \nodata           & 00 49 26.82 &\m30 56 18.2 & SAB(r)ab                   &15529 & 120 & 14909 & 16151 & 0.13  & 0.17   & 1995  \\
IRAS\,F00521-7054   & \nodata           & 00 52 06.38 &\m70 54 18.9 & E/S0 Sy2                   &20656 & 30  & 20344 & 21629 & 0.063 & 0.10   & 1995  \\
NGC\,300\tf         & \nodata           & 00 52 31.75 &\m37 57 15.1 & SA(s)d                     & 144  & 1   &\m132  & 733   & 0.086 & 0.11   & 1993  \\
NGC\,334            & \nodata           & 00 56 27.83 &\m35 23 04.9 & (R$'$)SB(s)b pec:          & 9210 & 10  & 7574	 & 10275 & 0.096 & 0.13   & 1997  \\
IRAS\,F01063-8034   & ESO\,013-G012     & 01 06 21.00 &\m80 34 24.0 & Sa                         & 5045\tg& 26& 4465  & 5626  & 0.057 & 0.063  & 1995  \\
IRAS\,F01063-8034   & ESO\,013-G012     & 01 06 21.00 &\m80 34 24.0 & Sa                         & 4249\tg& 21& 3084  & 6181  & 0.25  & 0.26   & 1998  \\
IC\,1631            & IRAS\,F01065-4644 & 01 06 31.60 &\m46 44 31.9 & Sab pec Sy2                & 9246 & 49  & 8913  & 9594  & 0.049 & \nodata& 1996  \\
NGC\,424\tz         & \nodata           & 01 09 09.64 &\m38 20 56.8 & (R)SB(r)0/a  Sy2           & 3496 & 30  & 1583	 & 4825  & 0.09  & 0.15   & 1997  \\
IC\,1657\tz         & \nodata           & 01 11 46.57 &\m32 54 55.1 & (R$'$)SB(s)bc  Sy2         & 3552 & 10  & 1652	 & 4881  & 0.066 & 0.084  & 1997  \\
IRAS\,01196-3254    & \nodata           & 01 19 35.40 &\m32 54 25.0 & S0                         & 9300 & 95  & 8699  & 9892  & 0.068 & 0.074  & 1995  \\
NGC\,526            & \nodata           & 01 21 38.00 &\m35 19 42.0 & pair                       & 5762 & 52  & 5180  & 6346  & 0.051 & 0.063  & 1995  \\
NGC\,625            & \nodata           & 01 32 54.10 &\m41 41 34.2 & SB(s)m? sp                 & 405  & 5   &\m28   & 837   & 0.048 & \nodata& 1993  \\
IRAS\,F01363-4016   & ESO\,297-G018     & 01 36 26.68 &\m40 15 53.8 & Sa: sp Sy2                 & 7555 & 15  & 7226  & 7899  & 0.039 & \nodata& 1996  \\
MCG-01-05-031       & IRAS\,F01428-0404 & 01 42 53.50 &\m04 04 37.0 & SB(rs)bc pec: Sy2          & 5456 & 22  & 3535  & 6632  & 0.12  & 0.14   & 1998  \\ 
AM\,0142-435        & ESO\,245-G005     & 01 42 58.00 &\m43 50 54.0 & IB(s)m                     & 395  & 6   &\m37   & 828   & 0.096 & \nodata& 1993  \\
Fairall\,377        & ESO\,197-G027     & 02 09 01.25 &\m49 56 01.4 & S Sy2                      &14240 & 90  & 13625 & 14857 & 0.13  & 0.17   & 1995  \\
NGC\,1032           & \nodata           & 02 36 49.09 & +00 52 44.5 & S0/a                       & 2694 & 18  & 1112  & 4228  & 0.13  & 0.14   & 1998  \\ 
IC\,1858            & \nodata           & 02 47 01.96 &\m31 29 46.0 & SA0+:                      & 6070 & 15  & 4514  & 7611  & 0.12  & 0.14   & 1998  \\
NGC\,1125\tz        & \nodata           & 02 49 20.31 &\m16 51 19.7 & (R$'$)SAB(rl:)$0^+$ Sy2    & 3297 & 22  & 1381	 & 4625  & 0.093 & 0.16   & 1997  \\
MCG\,-02-08-039     & \nodata           & 02 58 05.70 &\m11 36 50.0 & SAB(rs)a pec: Sy2          & 8874 & 90  & 7270	 & 9969  & 0.066 & 0.075  & 1997  \\
NGC\,1194           & \nodata           & 03 01 16.45 &\m01 17 53.8 & SA0+: Sy1                  & 4015 & 30  & 3695  & 4353  & 0.045 & \nodata& 1996  \\
NGC\,1209           & \nodata           & 03 03 42.80 &\m15 48 14.0 & E6:                        & 2600 & 18  & 2254  & 2988  & 0.045 & \nodata& 1996  \\
IRAS\,F03106-0254   & \nodata           & 03 10 37.49 &\m02 54 28.0 & S0 Sy2                     & 8154 & 90  & 7563  & 8747  & 0.086 & 0.091  & 1995  \\
IRAS\,03125+0119    & \nodata           & 03 12 30.35 & +01 19 25.5 & Sy2                        & 7200 & 70  & 6612  & 7789  & 0.095 & 0.11\ti& 1995  \\
NGC\,1313           & \nodata           & 03 17 39.00 &\m66 40 42.0 & SB(s)d HII                 & 475  & 3   & 25    & 890   & 0.062 & \nodata& 1993  \\
IRAS\,F03278-4329   & \nodata           & 03 27 48.69 &\m43 29 24.6 & Sy2                        &17508 & 90  & 16880 & 18138 & 0.13  & 0.17   & 1995  \\
NGC\,1365\tg        & \nodata           & 03 31 41.80 &\m36 18 26.6 & (R$'$)SBb(s)b  Sy1.8       & 1636 & 1   &\m228	 & 2982  & 0.084 & 0.08   & 1997  \\
NGC\,1448           & \nodata           & 03 42 52.70 &\m44 48 05.0 & SAcd: sp                   & 1164 & 5   & 730   & 1599  & 0.10  & \nodata& 1993  \\
\\
\\
\\
\\
ESO\,549-G040       & \nodata           & 03 54 56.09 &\m18 55 16.3 & Sb-c                       & 7534 & 10  & 5970	 & 8655  & 0.096 & 0.12   & 1997  \\
3C98                & \nodata           & 03 56 10.20 & +10 17 32.0 & E1?                        & 9130 & 42  & 7376	 & 10075 & 0.060 & 0.072  & 1997  \\
IRAS\,F04023-1638   & \nodata           & 04 02 21.5  &\m16 38 27   & \nodata                    & 8694 & 150 & 8686  &10286  & 0.17  & 0.18\ti & 1998  \\
NGC\,1587           & II\,Zw\,012       & 04 28 05.50 & +00 33 16.9 & E pec                      & 3694 & 12  & 3332  & 4071  & 0.035 & \nodata& 1996  \\
ESO\,552-G004       & \nodata           & 04 45 17.08 &\m17 41 05.1 & SA(rl)0\^{ }0              & 9007 & 20  & 7376	 & 10075 & 0.058 & 0.099  & 1997  \\
ESO\,485-G016       & \nodata           & 04 46 50.31 &\m23 49 00.8 & SAB(rs)ab:                 & 8129 & 22  & 6503	 & 9174  & 0.07  & 0.12   & 1997  \\
NGC\,1684           & \nodata           & 04 50 00.80 &\m03 11 20.0 & E+ pec:                    & 4456 & 25  & 2933  & 6085  & 0.13  & 0.14   & 1998  \\
NGC\,1705           & \nodata           & 04 53 06.20 &\m53 26 27.0 & SA0- pec: HII              & 628  & 9   & 240   & 1107  & 0.10  & \nodata& 1993  \\
NGC\,1808           & \nodata           & 05 05 58.58 &\m37 34 36.5 & (R$'$)SAB(s:)b Sy2         & 1000 & 5   & 566   & 1435  & 0.087 & \nodata& 1993  \\
NGC\,1808\tg        & \nodata           & 05 05 58.58 &\m37 34 36.5 & (R$'$)SAB(s:)b Sy2         & 1000 & 5   &\m1001	& 2445  & 0.078 & 0.11   & 1997  \\
IRAS\,F05189-2524   & \nodata           & 05 18 58.90 &\m25 24 39.0 & pec Sy2                    &12760 & 54  & 12150 & 13370 & 0.11  & 0.11   & 1995  \\
ESO\,253-G003       & \nodata           & 05 23 53.00 &\m46 02 54.0 & Sa?    Sy2                 & 12747& 21  & 10589	& 13280 & 0.09  & 0.10   & 1997  \\
IRAS\,F05238-4602   & ESO\,253-G003     & 05 23 53.00 &\m46 02 54.0 & Sa? Sy2                    &12738 & 22  & 12099 & 13319 & 0.11  & 0.11   & 1995  \\
IC\,422             & \nodata           & 05 30 05.15 &\m17 15 34.5 & \nodata                    & 2698 & 150 & 1218  & 4294  & 0.13  & 0.11   & 1998  \\
EXO\,0556.3-3820    & IRAS\,F05563-3820 & 05 56 21.10 &\m38 20 17.1 & Sy1                        &10154 & 180 & 9728  & 10929 & 0.068 & 0.08   & 1995  \\
NGC\,2207           & \nodata           & 06 14 14.40 &\m21 21 14.0 & SAB(rs)bc pec              & 2741 & 15  & 2361  & 3240  & 0.11\ti&\nodata& 1993  \\
ESO\,491-G021       & \nodata           & 07 07 49.00 &\m27 29 36.0 & SB(r)ab? pec               & 2847 & 68  & 2361  & 3240  & 0.11\ti&\nodata& 1993  \\
NGC\,2369           & \nodata           & 07 16 05.00 &\m62 15 12.0 & (R$'$\_2)SAB(s)ab          & 3237 & 36  & 2660  & 3541  & 0.10\ti&\nodata& 1993  \\
NGC\,2442           & \nodata           & 07 36 33.13 &\m69 24 58.3 & SAB(s)bc pec               & 1449 & 7   & 863   & 1997  & 0.089 & 0.089  & 1995  \\
ESO\,495-G021       & He\,2-010         & 08 34 07.00 &\m26 14 06.0 & I0? pec starburst          & 873  & 1   & 440   & 1308  & 0.084 & \nodata& 1993  \\
NGC\,2623           & Arp\,243          & 08 35 25.18 & +25 55 50.7 & LINER, triple              & 5535 & 8   & 4954  & 6118  & 0.051 & 0.057  & 1995  \\
NGC\,2640           & \nodata           & 08 36 05.00 &\m54 56 51.0 & SA0$^-$                    & 1051 & 32  & 736   & 1381  & 0.11  & \nodata& 1996  \\
NGC\,2835           & \nodata           & 09 15 37.00 &\m22 08 42.0 & SAB(rs)c                   & 888  & 5   & 455   & 1323  & 0.086 & \nodata& 1993  \\
NGC\,2845           & \nodata           & 09 16 37.00 &\m37 48 00.0 & SA(rs)a Sy2                & 2530 & 6   & 2212  & 2863  &0.041\ti&\nodata& 1996  \\
IRAS\,F09182-0750   & MCG\,-01-24-012   & 09 18 18.58 &\m07 50 35.4 & SAB(rs)c: Sy2              & 5936 & 90  & 5353  & 6520  & 0.12  & 0.12   & 1995  \\ 
NGC\,2915           & \nodata           & 09 26 31.00 &\m76 24 30.0 & I0                         & 468  & 5   & 36    & 901   & 0.098 & \nodata& 1993  \\
NGC\,2997\tf        & \nodata           & 09 43 27.35 &\m30 57 32.8 & SA(s)c                     & 1087 & 4   & 653   & 1522  & 0.13  & \nodata& 1993  \\
ESO\,434-G040       & \nodata           & 09 45 28.43 &\m30 42 57.0 & (RL)SA(l)0\^0  Sy2          & 2482 & 42  & 590	  & 3806  & 0.093 & 0.10   & 1997  \\
ESO\,373-G29\tz     & \nodata           & 09 45 33.10 &\m32 36 18.0 & (R)SAB(r)a   Sy2           & 2802 & 10  & 886	  & 4125  & 0.058 & 0.090  & 1997  \\
NGC\,3059           & \nodata           & 09 49 38.00 &\m73 41 12.0 & SB(rs)c                    & 1260 & 6   & 826   & 1696  & 0.089\ti&\nodata& 1993 \\
NGC\,3078           & \nodata           & 09 56 08.10 &\m26 41 13.0 & E2-3                       & 2495 & 25  & 2136  & 2869  & 0.032 & \nodata& 1996  \\
NGC\,3109\tf        & \nodata           & 10 00 49.00 &\m25 55 00.0 & SB(s)m                     & 403  & 1   & 68    & 934   & 0.1   & 0.14   & 1993  \\
IC\,2545            & \nodata           & 10 03 53.00 &\m33 38 30.0 & \nodata                    & 10267& 86  & 8573	 & 11334 & 0.07  & 0.075  & 1997  \\
\\
\\
\\
\\
IC\,2554            & \nodata           & 10 07 30.20 &\m66 47 02.0 & SB(s)bc pec: pair?         & 1474 & 30  & 943   & 1814  & 0.12\ti&\nodata& 1993  \\
FAIRALL\,1149\tz    & \nodata           & 10 11 08.00 &\m35 44 06.0 & (R$'$\_2)SB(s)ab Sy2       & 8530 & 14  & 6897	 & 9592  & 0.058 & 0.093  & 1997  \\ 
NGC\,3175           & \nodata           & 10 12 25.00 &\m28 37 24.0 & SAB(s)b                    & 1101 & 5   & 661   & 1530  & 0.048\ti&\nodata& 1993 \\
ESO\,317-G023       & \nodata           & 10 22 31.00 &\m39 03 06.0 & (R$'$\_1)SB(rs)a           & 2892 & 19  & 2453  & 3332  & 0.085\ti&\nodata& 1993 \\
NGC\,3256           & \nodata           & 10 25 43.40 &\m43 38 48.0 & Pec merger starburst       & 2738 & 28  & 2299  & 3178  & 0.080\ti&\nodata& 1993 \\
NGC\,3281           & \nodata           & 10 29 35.90 &\m34 35 46.0 & SAB(rs+)a Sy2              & 3200 & 22  & 3059  & 3942  & 0.083\ti&\nodata& 1993 \\
NGC\,3281\tz        & \nodata           & 10 29 35.90 &\m34 35 46.0 & SAB(rs+)a Sy2              & 3200 & 22  & 1543	 & 4789  & 0.078 & 0.099  & 1997  \\
IRAS\,F10329-1352   & MCG-02-27-009     & 10 32 59.60 &\m13 52 14.0 & SB(rs)0+ pec? Sy2          & 4529 & 31  & 4165  & 4908  & 0.023 & \nodata& 1996  \\
MCG-02-27-009       & IRAS\,F10329-1351 & 10 32 59.60 &\m13 52 14.0 & SB(rs)0+ pec? Sy2          & 4529 & 31  & 3092  & 6189  & 0.12  & 0.15   & 1998  \\ 
NGC\,3511           & \nodata           & 11 00 57.00 &\m22 49 00.0 & SAB(s)c Sy1                & 1106 & 4   & 672   & 1541  & 0.088 & \nodata& 1993  \\
ESO\,438-G020       & \nodata           & 11 16 27.00 &\m29 09 06.0 & S? LINER                   & 9121 & 10  & 7435	 & 10075 & 0.087 & 0.12   & 1997  \\
IRAS\,F11186-0242   & \nodata           & 11 18 39.08 &\m02 42 36.5 & SAB(s)b HII Sy2            & 7464 & 23  & 6875  & 8054  & 0.051 &0.057\ti& 1995  \\
ESO\,320-G030       & Fairall\,1151     & 11 50 39.90 &\m38 51 07.0 & (R$'$\_1)SAB(r)a  NELG     & 3232 & 19  & 2660  & 3541  & 0.094 & \nodata& 1993  \\
NGC\,4507\tz        & \nodata           & 12 32 54.50 &\m39 38 02.0 & SAB(s)ab    Sy2            & 3538 & 9   & 1606	 & 4832  & 0.096 & 0.14   & 1997  \\
M\,104              & Sombrero          & 12 37 23.39 &\m11 20 54.9 & SA(s)a LINER Sy1.9         & 1024 & 5   &\m504  & 2596  & 0.13  & 0.15   & 1998  \\ 
IC\,3639\tz         & \nodata           & 12 38 10.60 &\m36 28 54.0 & SB(rs)bc:   Sy2            & 3275 & 7   & 1394	 & 4527  & 0.06  & 0.12   & 1997  \\
ESO\,172-G010\tz    & \nodata           & 12 44 46.3  &\m53 16 41   & SBm                        & 1829 & 30  &\m80	  & 3171  & 0.070 & 0.072  & 1997  \\
IRAS\,F12585-3208   & ESO\,443-G029     & 12 58 36.00 &\m32 07 54.0 & SA(r)c                     & 9397 & 24  & 7912	 & 9411  & 0.11  & 0.11   & 1997  \\
ESO\,508-G05\tz     & \nodata           & 13 04 14.00 &\m23 39 00.0 & SB(rl)0/a:                 & 2947 & 40  & 1052	 & 4253  & 0.081 & 0.11   & 1997  \\
NGC\,4968\tz        & \nodata           & 13 04 24.00 &\m23 24 42.0 & (R$'$)SAB0\^0 Sy2          & 2957 & 26  & 1052	 & 4243  & 0.13  & 0.14   & 1997  \\
IRAS\,F13109-1509   & \nodata           & 13 10 54.8  &\m15 10 00   & SAB(s)dm                   & 2502 & 6   & 910   & 4036  & 0.12  & 0.14   & 1998  \\
NGC\,5068           & \nodata           & 13 16 12.10 &\m20 46 36.0 & SB(s)d                     & 673  & 5   & 566   & 1435  & 0.082 & \nodata& 1993  \\
NGC\,5078           & \nodata           & 13 17 04.84 &\m27 08 50.4 & SA(s)a: sp                 & 2168 & 6   & 567   & 3660  & 0.19  & 0.29   & 1998  \\
IRAS\,F13174-1651   & VV802             & 13 17 25.9  &\m16 51 13   & pair                       & 6296 & 150 & 4704  & 7801  & 0.15  & 0.15   & 1998  \\
ESO\,324-G024       & UKS\,1324-412     & 13 24 42.00 &\m41 13 18.0 & IABm:                      & 513  & 6   & 81    & 947   & 0.12\ti&\nodata& 1993  \\
M\,83\tf            & \nodata           & 13 34 11.55 &\m29 36 42.2 & SAB(s)c  HII starburst     & 516  & 4   & 45    & 910   & 0.097 & 0.099  & 1993  \\
NGC\,5253           & \nodata           & 13 37 05.12 &\m31 23 13.2 & Im pec HII starburst       & 404  & 4   &\m28   & 837   & 0.14  & \nodata& 1993  \\
ESO\,221-IG010      & \nodata           & 13 47 48.00 &\m48 48 30.0 & pair                       & 3099 & 39  & 2588  & 3469  & 0.096\ti&\nodata& 1993 \\
NGC\,5495           & IRAS\,F14095-2652 & 14 09 31.00 &\m26 52 24.0 & (R$'$)SA(rs)b Sy2            & 6737 &  9  & 6370  & 7125  & 0.032 & \nodata& 1996  \\
UKS\,1424-460       & \nodata           & 14 24 48.00 &\m46 04 48.0 & IB(s)m                     & 397  & 68  &\m39   & 826   & 0.093 & \nodata& 1993  \\
NGC\,5643           & \nodata           & 14 29 27.30 &\m43 57 16.0 & SAB(rs)c Sy2               & 1199 & 5   & 666   & 1535  & 0.059 & \nodata& 1993  \\
NGC\,5643\tz        & \nodata           & 14 29 27.30 &\m43 57 16.0 & SAB(rs)c   Sy2             & 1199 & 5   &\m706	 & 2528  & 0.070 & 0.087  & 1997  \\
NGC\,5833           & \nodata           & 15 06 42.00 &\m72 40 12.0 & SAB(r)c                    & 3071 & 8   & 2590  & 3471  & 0.10\ti&\nodata& 1993  \\
\\
\\
\\
\\
IRAS\,F15366+0544   & ARK481            & 15 36 36.77 & +05 43 58.9 & E                          & 7781 & 42  & 6181  & 9283  & 0.15  & 0.17   & 1998  \\ 
IRAS\,15374-1817    & ESO\,583-G002     & 15 37 28.6  &\m18 16 55   & SB(rs)bc Sy1               & 7042 & 10  & 6724  & 7395  & 0.037 & \nodata& 1996  \\ 
IRAS\,15480-0344    & \nodata           & 15 48 03.98 &\m03 44 17.5 & S0 Sy2                     & 9084 & 90  & 8489  & 9680  & 0.068 &0.074\ti& 1995  \\
3C327               & IRAS\,F15599+0206 & 15 59 55.67 & +02 06 12.3 & Sy2                        &31170 & 150 & 30495 & 31863 & 0.044 & 0.063  & 1995  \\
NGC\,6156           & \nodata           & 16 30 28.60 &\m60 30 53.0 & (R$'$)SAB(rs)c             & 3300 & 30  & 2660  & 3541  & 0.069\ti&\nodata& 1993 \\
ESO\,137-G034       & \nodata           & 16 31 01.00 &\m57 58 30.0 & SAB(s)0/a? Sy2             & 2747 & 16  & 2411  & 3290  & 0.093\ti&\nodata& 1993 \\
NGC\,6215           & \nodata           & 16 46 47.43 &\m58 54 26.5 & SA(s)c pec                 & 1560 & 6   & 915   & 1786  & 0.079 & \nodata& 1993  \\
NGC\,6221           & \nodata           & 16 48 25.40 &\m59 08 00.0 & SB(s)bc pec Sy2            & 1482 & 6   & 915   & 1786  & 0.079 & \nodata& 1993  \\
NGC\,6300           & \nodata           & 17 12 18.00 &\m62 45 54.0 & SB(rs)b Sy2                & 1110 & 6   & 676   & 1545  & 0.059 & \nodata& 1993  \\
IC\,4662            & He\,2-269         & 17 42 12.00 &\m64 37 18.0 & IBm                        & 308  & 4   &\m125  & 740   & 0.077 & \nodata& 1993  \\
IRAS\,F18325-5926   & Fairall\,49       & 18 32 32.49 &\m59 26 40.2 & Sa Sy2                     & 6065 & 70  & 5481  & 6650  & 0.046 & 0.051  & 1995  \\
H1846-786           & IRAS\,F18389-7834 & 18 39 03.5  &\m78 35 06   & Sy1                        &22275 & 15  & 22135 & 23435 & 0.046 & 0.074  & 1995  \\
IC\,4778            & \nodata           & 18 45 25.00 &\m61 46 36.0 & (R$'$)SB(s)a               & 4971 & 45  & 3388  & 6481  & 0.13  & 0.15   & 1998  \\
NGC\,6753           & \nodata           & 19 07 11.00 &\m57 07 54.0 & (R$¹$)SA(r)b               & 3124 & 26  & 2660  & 3541  & 0.098 & \nodata& 1993  \\
IRAS\,F19254-7245   & Super~Antenna     & 19 25 29.86 &\m72 45 37.5 & pair                       &18500 & 80  & 17868 & 19133 & 0.091 & 0.11   & 1995  \\
IRAS\,F19254-7245   & Super~Antenna     & 19 25 29.86 &\m72 45 37.5 & pair                       & 18500& 80  & 16926 & 20018 & 0.15  & 0.15   & 1998  \\
IRAS\,F19254-7245   & Super~Antenna     & 19 25 29.86 &\m72 45 37.5 & pair                       & 18500& 80  & 17704	& 19299 & 0.072 & 0.075  & 1997  \\
IC\,4859            & \nodata           & 19 25 52.00 &\m66 25 18.0 & SA(s)bc                    & 5003 &     & 4424  & 5584  & 0.063 & 0.076  & 1995  \\
NGC\,6810           & IRAS\,F19393-5846 & 19 39 21.00 &\m58 46 30   & SA(s)ab:sp Sy2             & 2031 & 10  & 1714  & 2363  & 0.043 & \nodata& 1996  \\
NGC\,6810           & \nodata           & 19 39 21.00 &\m58 46 30.0 & SA(s)ab:sp Sy2             & 2031 & 10  & 383   & 3471  & 0.12  & 0.12   & 1998  \\ 
IRAS\,F19395-7000   & ESO\,073-G005     & 19 39 35.00 &\m70 00 12.0 & SAa?                       & 3786 &     & 2215  & 5304  & 0.13  & 0.13   & 1998  \\
NGC\,6868           & \nodata           & 20 06 16.40 &\m48 31 39.0 & E2                         & 2854 & 15  & 2638  & 3346  & 0.076 & 0.076  & 1995  \\
IC\,5052            & \nodata           & 20 47 22.00 &\m69 23 30.0 & SBd: sp                    & 598  & 5   & 165   & 1032  & 0.11  & \nodata& 1993  \\
ESO\,286-G018\tj    & IRAS\,F20545-4334 & 20 54 30.30 &\m43 34 10.2 & SB(s)bc? Sp                &9147  & 10  & 8552  & 9744\tk & 0.11  & 0.13\ti& 1995  \\
NGC\,6987           & \nodata           & 20 54 41.55 &\m48 49 25.0 & E0:                        & 5239 & 27  & 4889  & 5636  & 0.023 & \nodata& 1996  \\
IRAS\,F20559-5251   & ESO\,187-G042     & 20 55 54.65 &\m52 51 33.1 & (R$'$)SB(s)a Sy2             & 7180 & 36  & 6852  & 7523  & 0.038 & \nodata& 1996  \\
Mrk\,897            & \nodata           & 21 05 15.07 & +03 40 31.9  & Sy2                        & 7897 &  1  & 7176	 & 8667  & 0.066 & 0.070  & 1997  \\
IRAS\,F21497-0824   & \nodata           & 21 49 47.20 &\m08 24 31.8 & \nodata                    & 10330& 41  & 7952  & 11871 & 0.15  & 0.17   & 1998  \\
IRAS\,F21529-6955   & ESO\,075-G041     & 21 52 58.00 &\m69 55 41.2 & SA0- Radio gal Sy2         & 8476 & 31  & 6907  & 9168  & 0.15  & 0.17   & 1998  \\
IC\,5152\tf         & \nodata           & 21 59 26.58 &\m51 32 14.5 & IA(s)m                     & 124  & 3   &\m157  & 708   & 0.099 & 0.046  & 1993  \\
NGC\,7205           & \nodata           & 22 05 10.80 &\m57 41 16.0 & SA(s)bc LINER              & 1690 & 7   & 915   & 1786  & 0.08  & \nodata& 1993  \\
NGC\,7213           & \nodata           & 22 06 08.40 &\m47 24 45.0 & SA(s)0\^0 Sy1.5            & 1792 & 9   & 1225  & 2361  & 0.095 & 0.10   & 1995  \\
3C445               & \nodata           & 22 21 14.72 &\m02 21 25.3 & N BLRG Sy1                 &16848 & 15  & 16223 & 17475 & 0.12  & 0.14   & 1995  \\
\\
\\
\\
\\
NGC\,7410\tz        & \nodata           & 22 52 09.74 &\m39 55 45.0 & SB(s)a LINER Sy2           & 1751 & 28  &\m158	 & 3071  & 0.099 & 0.12   & 1997  \\
NGC\,7410           & \nodata           & 22 52 11.00 &\m39 55 42.0  & SB(s)a LINER Sy2           & 1751 & 28  & 166   & 1798  & 0.13  & 0.14   & 1998  \\ 
IC\,1459            & IRAS\,F22544-3643 & 22 54 23.11 &\m36 43 47.4 & E3                         & 1691 & 18  & 138   & 3236  & 0.13  & 0.14   & 1998  \\
IC\,1459            & \nodata           & 22 54 23.11 &\m36 43 47.4 & E3                         & 1691 & 18  & 1374  & 2022  &0.068\ti&\nodata& 1996  \\
IRAS\,F23031-3052   & ESO\,469-G011     & 23 03 05.60 &\m30 52 55.0 & pec Sy2                    & 8504 & 13  & 8274  & 8951  & 0.043 & \nodata& 1996  \\
ESO\,407-G018       & UKS\,2323-326     & 23 23 47.29 &\m32 39 50.4 & IB(s)m pec:                & 62   & 5   &\m369  & 494   & 0.055 & \nodata& 1993  \\
IRAS\,F23377-4447   & ESO\,292-G009     & 23 37 44.45 &\m44 47 30.9 & SAB(r)b Sy                 &15444 & 20  & 14825 & 16066 & 0.13  & 0.18   & 1995  \\
ESO\,471-G006       & AM\,2341-321      & 23 41 08.53 &\m32 14 01.5 & SB(s)m: sp                 & 267  & 8   &\m105  & 760   & 0.076 & \nodata& 1993  \\
NGC\,7793\tf        & \nodata           & 23 55 15.50 &\m32 52 03.0 & SA(s)d                     & 230  & 2   &\m132  & 733   & 0.077 & 0.084  & 1993  \\
IRAS\,F23565-7631   & \nodata           & 23 56 30.5  &\m76 31 19   & \nodata                    & 25183& 150 & 23625 & 26715 & 0.21  & 0.29   & 1998  \\

\enddata

\tablenotetext{(a)}{Catalog coordinates listed by NED.  Pointing positions differed by $>10\%$ of the $1\rlap{.}'4$ 
beam (half-power full-width) in 10 cases--
                    IRAS\,F09182-0750 ($13\rlap{.}''1$),
                    IRAS\,F13109-1509 ($13\rlap{.}''1$), 
                    IRAS\,F23377-4447 ($8\rlap{.}''4$),
                    NGC\,625 ($24\rlap{.}''0$),
                    NGC\,3621 ($11\rlap{.}''1$),
                    NGC\,5128 ($12\rlap{.}''6$),
                    NGC\,5253 ($17\rlap{.}''8$),
                    NGC\,5643 ($8\rlap{.}''5$),
                    NGC\,7410 ($9\rlap{.}''5$), and
                    NGC\,7793 ($9\rlap{.}''5$). }

\tablenotetext{(b)}{Optical morphology as listed by NED,
                    largely following the notation of \citet{devaucouleurs76}.}

\tablenotetext{(c)}{Heliocentric velocity referenced by NED, assuming the optical definition of Doppler shift.}

\tablenotetext{(d)}{Velocities corresponding to the highest and lowest observed frequencies.  Excludes
                    guard bands of 2 MHz in 1997 and 1998, and 5 MHz in 1996, in which the effects of 
                    filter roll-off can be severe.}

\tablenotetext{(e)}{RMS noise level in a 0.84\kms~spectral channel after Hanning smoothing.}
                    
\tablenotetext{(f)}{Multiple fields observed: 10 in NGC\,55, 5 in NGC\,300, 2 in NGC\,2997, 8 in NGC\,3109, 
36 in M\,83, 14 in NGC\,7793, and 6 in IC\,5152.}

\tablenotetext{(g)}{Estimates of systemic velocity in the literature disagree. See review in $\S3.2$. }

\tablenotetext{(h)}{Object previously observed with a roughly 600\kms~bandwidth centered on the systemic 
velocity. No emission reported.  Observations reported here cover velocities red and blueshifted outside that band.
Quoted velocity range reflects composite of old and new observations.}

\tablenotetext{(i)}{Reduced pointing accuracy due to high winds or attentuation due to overcast may
                    have degraded sensitivity.}

\tablenotetext{(j)}{Paired with ESO\,286-G017 at $9778\pm5$\kms, which is type Sy2.}

\tablenotetext{(k)}{Velocities between 9408 and 9475\kms~not observed.}

\end{deluxetable}

\clearpage

\begin{deluxetable}{llr}
\tabletypesize{\footnotesize}
\tablenum{3}
\tablewidth{4.0in}
\tablecolumns{3}
\tablecaption{Properties of IRAS\,F01063-8034}
\tablehead{ \multicolumn{2}{c}{Measurement} & \colhead{Ref.} }
\startdata

Alias                           & ESO\,013-G012  \\
Hubble Type                     & Sa                                   &    \\
Optical $\alpha_{2000}$         & $\phn\,01^h07^m01^s  \pm 3'' $       & 1  \\
~~~~~~~~~~~$\delta_{2000}$      & $-80^\circ18'24''    \pm 8'' $       &    \\
$\lambda 6$cm~~~$\alpha_{2000}$ & $\phn\,01^h07^m01\rlap{.}^s70\pm 0\rlap{.}^s4$ & 2  \\
 ~~~~~~~~~      $\delta_{2000}$ & $-80^\circ18'28\rlap{.}''1\pm 1''$   &    \\ 
Position Angle                  & $153^\circ\pm2^\circ$\tablenotemark{(a)}   &    \\
Inclination                     & $83^\circ$     \tablenotemark{(b)}   &    \\
$V_{\rm sys}$                   & $5047\pm 21$\kms\tablenotemark{(c)}  & 3  \\
                                & $4249\pm 27$\kms                     & 4  \\
                                & $4285\pm 35$\kms                     & 5  \\
$F_\nu$ (\phn12 $\mu$m)         & $0.16\pm0.02$ Jy                     & 6  \\
$F_\nu$ (\phn25 $\mu$m)         & $0.14\pm0.02$ Jy                     & 6  \\
$F_\nu$ (\phn60 $\mu$m)         & $1.52\pm0.06$ Jy                     & 6  \\
$F_\nu$ (100 $\mu$m)            & $6.51\pm0.3$  Jy                     & 6  \\
$F_\nu$ ($\lambda6.3$ cm)       & $13.4\pm0.4$ mJy~~\tablenotemark{(d)}& 2  \\
$F_\nu$ ($\lambda3.5$ cm)       & $11.7\pm0.4$ mJy~~\tablenotemark{(d)}& 2  \\
$F_\nu$ ($\lambda13 $ cm)       & $<4$ mJy~~\tablenotemark{(e)}        & 7  \\

\tablenotetext{(a)}{Position angle estimated from Figure\,1.}
\tablenotetext{(b)}{Inclination estimated from the axial ratio of isophotes
\citep{vdb88}.}

\tablenotetext{(c)}{Systemic heliocentric velocities, assuming the optical definition of
Doppler shift.}

\tablenotetext{(d)}{Integrated flux density. Beam half-power full width 1 - $2''$.}

\tablenotetext{(e)}{Fringe spacing of interferometric observation $0\rlap{.}''09$.}

\tablerefs{1: ESO/Uppsala Survey, 2:  McGregor\etal, in prep., 3: \citet{sadler84}, 4:
\citet{dacosta91}, 5: this work, 6: \citet{moshir92}, 7: \citet{slee94}. }

\enddata
\end{deluxetable}

\end{document}